\documentclass[aps,twocolumn,showpacs,preprintnumbers,superscriptaddress,amsmath,amssymb]{revtex4}

    \usepackage[dvips]{graphicx} 

\usepackage[english]{babel}
\usepackage{blindtext}
\usepackage{dcolumn}
\usepackage{bm}
\usepackage{ulem}
\usepackage[dvips]{color} 

\newcommand{\ie}{{\it i.e. }}
\newcommand{\eg}{{\it e.g. }}

\newcommand{\co}[2]{\ifcase #1 \or #2 \fi}

\newcommand{\bscco}{Bi$_{2}$Sr$_{2}$CaCu$_{2}$O$_{8}$\,}

\newcommand{\celsius}{\,$^\circ$C}

\newcommand\degrees[1]{\ensuremath{#1^\circ}}

\newif\ifnote

\begin{document}


\title{Coherent THz emission of
intrinsic Josephson junction stacks in the hot spot regime}

\author{H.B.~Wang}
\affiliation{National Institute for Materials Science, Tsukuba 3050047, Japan}
\author{S.~Gu\'{e}non}
\author{B.~Gross}
\affiliation{
Physikalisches Institut --
Experimentalphysik II
and
Center for Collective Quantum Phenomena,
Universit\"{a}t T\"{u}bingen,
Auf der Morgenstelle 14,
D-72076 T\"{u}bingen,
Germany
}
\author{J.~Yuan}
\affiliation{National Institute for Materials Science, Tsukuba 3050047, Japan}
\author{Z.G.~Jiang}
\author{Y.Y.~Zhong}
\affiliation{Research Institute of Superconductor Electronics, Nanjing University, Nanjing 210093, China}
\author{M.~Gruenzweig}
\affiliation{
Physikalisches Institut --
Experimentalphysik II
and
Center for Collective Quantum Phenomena,
Universit\"{a}t T\"{u}bingen,
Auf der Morgenstelle 14,
D-72076 T\"{u}bingen,
Germany
}
\author{A.~Iishi}
\affiliation{National Institute for Materials Science, Tsukuba 3050047, Japan}
\author{P.H.~Wu}
\affiliation{Research Institute of Superconductor Electronics, Nanjing University, Nanjing 210093, China}
\author{T.~Hatano}
\affiliation{National Institute for Materials Science, Tsukuba 3050047, Japan}
\author{D.~Koelle}
\affiliation{ Physikalisches Institut -- Experimentalphysik II and Center for Collective
Quantum Phenomena, Universit\"{a}t T\"{u}bingen, Auf der Morgenstelle 14, D-72076
T\"{u}bingen, Germany }
\author{R.~Kleiner}
\affiliation{
Physikalisches Institut --
Experimentalphysik II
and
Center for Collective Quantum Phenomena,
Universit\"{a}t T\"{u}bingen,
Auf der Morgenstelle 14,
D-72076 T\"{u}bingen,
Germany
}%

\date{\today}

\begin{abstract}
We report on THz emission measurements and low temperature scanning laser imaging of
Bi$_{2}$Sr$_{2}$CaCu$_{2}$O$_{8}$
intrinsic Josephson junction stacks.
Coherent emission is observed at large dc input power, where a hot spot and a standing wave, formed in the ``cold'' part of the stack, coexist.
By varying the hot spot size the cavity resonance frequency and the emitted radiation can be tuned. The linewidth of radiation is much smaller than expected from the quality factor of the cavity mode excited. Thus, an additional mechanism of synchronization seems to play a role, possibly  arising from nonequilibrium processes at the hot spot edge.

\end{abstract}



\maketitle
Phase synchronization is one of the prerequisites to use Josephson junction arrays as tunable high frequency sources \cite{Darula99}. While Nb based junctions are limited to frequencies well below 1 THz, intrinsic Josephson junctions (IJJs)\cite{Kleiner92,Yurgens00} in \bscco (BSCCO) are, at least in principle, able to operate up to several THz. Stacks of many junctions can be made \eg by patterning mesa structures on top of single crystals.
For many years, investigations focused on small structures consisting of some 10 IJJs, with lateral mesa sizes of a few $\,\mu$m. Here, with few exceptions \cite{Clauss04,Katterwe09}, the IJJs in the stack tended to oscillate out-of-phase or were not synchronized at all.
Experimental and theoretical studies included
the generation of synchronous Josephson oscillations by moving Josephson vortices \cite{Kleiner94, Ustinov98,
Machida00, Bae07, Clauss04, Wang06,Katterwe09},
the use of shunting elements \cite{Wang00, Madsen04, Grib06, Grib09}, the excitation of Josephson
plasma oscillations via heavy quasiparticle injection \cite{Lee00,
Kume99} or the investigation of stimulated emission due to quantum
cascade processes \cite{Krasnov06}.
High-frequency emission of unsynchronized intrinsic junctions has
been observed up to 0.5\,THz \cite{Batov06}.

Recently, coherent off-chip THz radiation with an extrapolated output power of some
$\mu$W was observed from stacks of more than 600 IJJs, with lateral dimensions in the 100$\,\mu$m range  \cite{Ozyuzer07,Kadowaki08,Ozyuzer09a,Minami09,Kadowaki09}.
Phase synchronization involved a
cavity resonance oscillating along the short side of the mesa. This radiation was studied theoretically in a series of recent papers, either based on vortex-type or plasmonic excitations
\cite{Bulaevskii07, Koshelev08, Koshelev08b, Lin08,Hu09,Klemm09,Nonomura09,Tachiki09,Koyama09,Savelev10}
, or on nonequilibrium effects caused by quasiparticle injection \cite{Krasnov09}.

While in experiments \cite{Ozyuzer07,Kadowaki08,Ozyuzer09a,Minami09,Kadowaki09} THz emission was obtained at relatively low bias currents and moderate dc power input, using low temperature scanning laser microscopy (LTSLM) we have shown that standing wave patterns, presumably associated with THz radiation, can be obtained at high input power where, in addition, a hot spot (\ie a region heated to above the critical temperature $T_c$) forms within the mesa structure\cite{Wang09a}. Hot spot and waves seem to be correlated. The purpose of the present work is to investigate THz emission in this high power regime in detail, combining THz emission measurements and LTSLM. Apart from further clarifying the role of the hotspot, we are specifically interested in the question whether \textit{coherent} radiation can be achieved in spite of the high temperatures involved.

For the experiments BSCCO single crystals were grown by the floating zone technique in a four lamp arc-imaging
furnace. Below we discuss results from two samples. Sample 1 was patterned on a crystal that was annealed in vacuum at 650\celsius\ for 65 hours. It had a $T_c$ of 86.6 K and a transition width $\Delta T_c$ of 1.5 K. Sample 2 was made on a crystal ($T_c =$ 87.6 K, $\Delta T_c = 1.5$ K) annealed in vacuum at 600\celsius\ for 72 hours.
To provide good electrical contact the single crystals were cleaved
in vacuum and a 30\,nm Au layer was evaporated.
Then, conventional photolithography was used to define the mesa size in the $a$-$b$ plane (length 330$\,\mu$m, width 50$\,\mu$m  for both samples). Ar ion milling yielded mesas with measured thicknesses of, respectively 1$\,\mu$m (sample 1) and 0.7 $\,\mu$m (sample 2) along the
$c$-axis (corresponding to, respectively, stacks of 670 and 470 IJJs).
Insulating polyimide was used to surround the mesa
edge at which a Au wire was attached to the mesa by
silver paste.
Other Au wires were connected to the big single crystal
pedestal as grounds.
In order to provide a load line for stable operation, the mesas were
biased using a current source and variable resistor in parallel to
the mesa, cf.~Fig. 1 in Ref. \cite{Wang09a}.
The voltage measured across the mesa includes the resistance of the contacting Au wires and the resistance between these wires and the mesa ($4 \Omega$ for sample 1,  $6.5 \Omega$ for sample 2). In the data discussed below this resistance is subtracted.

\begin{figure}[tb]
\begin{center}
\includegraphics[width=\columnwidth,clip]{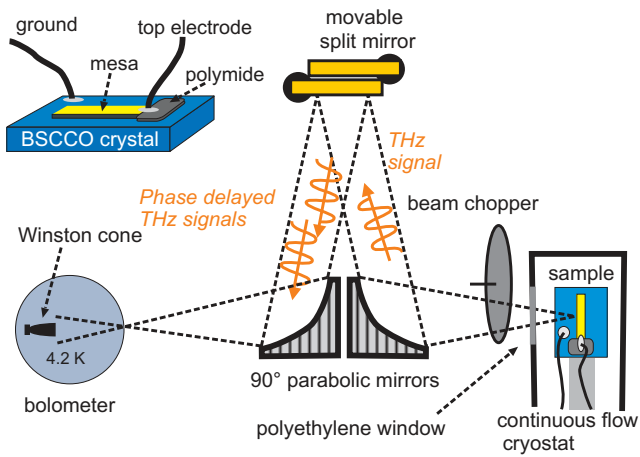}
\end{center}
\caption{(color online).
Schematic of the THz emission setup together with a sketch of the sample.
}
\label{fig:1}
\end{figure}

THz emission measurements were performed in Tsukuba; the samples were subsequently shipped to T\"ubingen for LTSLM.
In total we detected THz emission  from  7 (out of a total of 12) mesas on 5 different crystals. The LTSLM setup is described in Ref.\cite{Wang09a}. In brief, the beam of a diode laser (modulated at 10-80 kHz, spot size 1-2 $\,\mu$m) is deflected by a scanning unit and focused onto the sample surface. Local heating by 2--3 K in an area of a few $\mu$m$^2$ and about 0.5 $\mu$m in depth causes a change $\Delta V$ (detected using lock-in techniques) of the voltage $V$ across the mesa  serving as the contrast for the LTSLM image. Standing waves can be imaged due to the beam-induced local change of the quality factor, leading to a strong signal $\Delta V$ at antinodes and a weak signal at nodes. The edge of a hot spot leads to a strong signal, while $\Delta V$ is small in the interior of the hot spot and in the ''cold'' part of the mesa \cite{Wang09a}.
The THz emission setup is shown schematically in Fig.\ref{fig:1} together with a sketch of the sample. The interferometer is similar to the one in Ref. \onlinecite{Eisele07}.
The crystal is mounted in a continuous flow cryostat with a
polyethylene window. The emitted radiation, chopped with a frequency
of 13 Hz, is collected by a \degrees{90} off-axis parabolic mirror
(2'' diameter, focal length 100 mm) and deflected towards two
lamellar split mirrors dividing the incoming radiation into two
beams with almost the same intensity. A phase difference is
introduced by moving one of these mirrors with a translation stage.
Via a second parabolic mirror the radiation is deflected to the
bolometer (gauged by using a Gunn oscillator for reference). A 1 THz
cut-off type filter is used in front of the bolometer. The numbers
we quote below are based on the \textit{detected} power, ignoring
additional losses (by a factor of 2--5) from the continuous flow
cryostat window and the mirrors. The solid angle of our setup,
defined by the aperture of the Winston cone in front of the
bolometer, is 0.04 sr. The frequency resolution, depending on the
maximum mirror displacement, is 10--15 GHz.

%

\begin{figure}[tb]
\includegraphics[width=\columnwidth,clip]{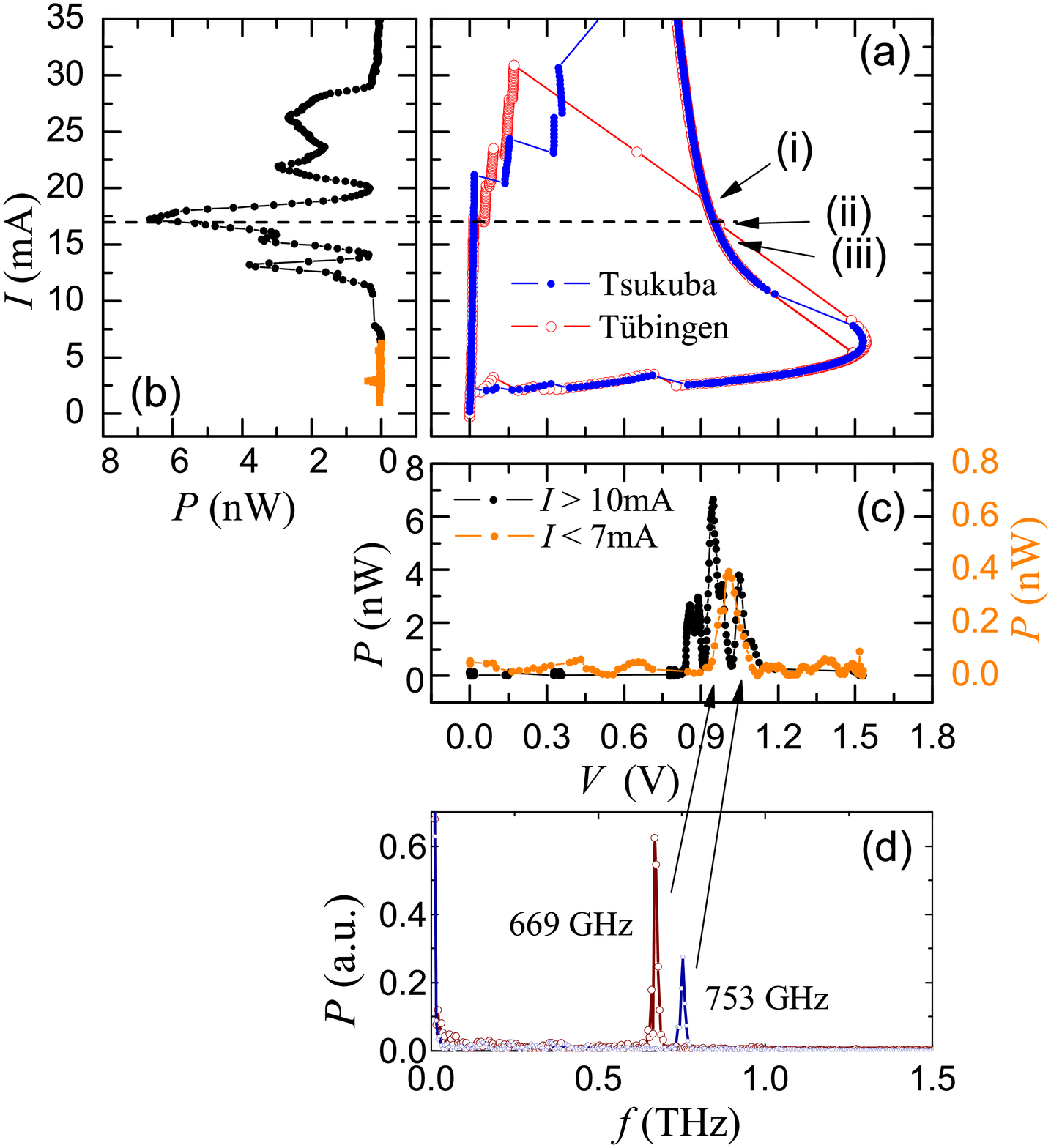}
\caption{(color online).
IVCs (a) of sample 1 at $T$ = 20K, as measured in Tsukuba [open (red) circles] and in T\"ubingen [solid (blue) circles]. Labelled arrows refer to bias values where LTSLM images are taken (c.f. Fig. \ref{fig:3}). Emission power detected by the bolometer as a function of current (b) and voltage (c). Black lines refer to high bias ($I >$10 mA), grey (orange) lines to low bias ($I <$7 mA). Arrows in (c) indicate where Fourier spectra (d) have been taken; the peak at 0.669 THz corresponds to $V = $ 944 mV and $I=$17.2 mA, the peak at 0.753 THz to $V = 1046$ mV and $I = $13.2 mA.
}
\label{fig:2}
\end{figure}
%

We first discuss THz emission data of sample 1, measured at $T=$20 K. Figure \ref{fig:2} (a) shows by  open circles the current voltage characteristic (IVC), as measured in Tsukuba. Starting from zero current, groups of junctions become resistive for $I>20$mA. At $I=30$ mA  the whole stack switches to the resistive state. Lowering the bias in this state the IVC is continuous down to about 10 mA where another switch to a current of about 8 mA occurs. As revealed by LTSLM this switch corresponds to the disappearance of the hot spot present at high bias. For 3 mA $< I <$ 10 mA the IVC is continuous until further switches and hystereses appear due to the retrapping to the zero voltage state of some of the IJJs in the stack. Figures \ref{fig:2} (b) and (c) show the emission power detected by the bolometer as a function of, respectively, $I$ and $V$. The largest response occurs in the high power (hot spot) regime and covers a current range from 10 mA up to about 30 mA. Here, the voltage across the mesa ranges from 0.83 to 1.13 V. If all $N$ IJJs in the stack are \textit{frequency} locked and participate in radiation we expect that the Josephson relation $V/N =\Phi_0f$ holds, where $\Phi_0$ is the flux quantum and $f$ is the emission frequency. The Fourier spectrum, at given bias, revealed a single sharp emission line (with a resolution limited $\Delta f \approx$ 12 GHz linewidth) continuously shifting with $V$. Two spectra are shown in Fig. \ref{fig:2} (d). From these data we infer $N = 676 \pm 5$ in very good agreement with the 1$\,\mu$m thickness of the mesa. The maximum detected power was 6 nW, corresponding to roughly 2 $\,\mu$W when extrapolating to 4$\pi$, even without taking additional losses into account. Thus the emitted power is comparable to the one quoted in Ref. \onlinecite{Ozyuzer07}.  Sample 1 also showed a bolometric response at low bias
[c.f. grey (orange) lines in Figs. \ref{fig:2} (b) and (c)]. Here, the detected power (of order 0.4 nW) was too low for Fourier spectrometry.

\begin{figure}
\includegraphics[width=\columnwidth,clip]{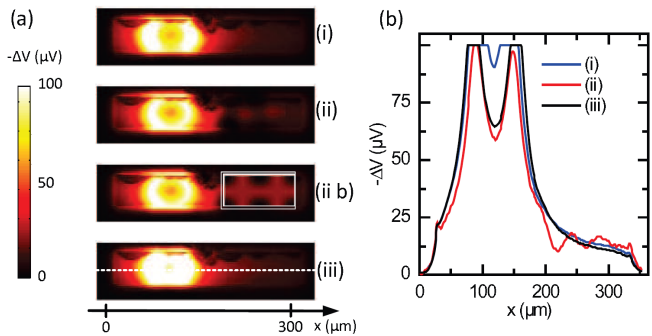}
\caption{(color online)
%
LTSLM imabges (a) and linescans along the center of the mesa (b) of sample 1 at T = 20 K, corresponding to the bias points (i),(ii) and (iii) indicated in Fig. \ref{fig:2}(a). In (a), image (ii b) shows a simulated standing wave pattern superposed on the data of image (ii).
}
\label{fig:3}
\end{figure}

The different peaks visible in the radiometer data of Figs.  \ref{fig:2} (b) and (c) are likely to correspond to different cavity modes excited in the mesa (a similar case is shown in Fig. 3 of Ref. \cite{Wang09a}). For the present sample, only for a bias near the main emission peak the standing wave pattern appeared clear enough to be analyzable.
Fig. \ref{fig:3}(a) shows LTSLM images at the bias points (i), (ii) and (iii) indicated on the IVC shown in Fig.  \ref{fig:2} (a) by solid circles. Fig. \ref{fig:3} (b) shows the corresponding linescans, taken along the center of the mesa. The bright feature visible in the left part of the mesa is the hot spot growing with increasing input power.  Image (ii) shows a standing wave pattern located in the right part of the mesa. There are 3 half waves along the long side of the mesa and one half wave along the short side. A simulation of this wave pattern, superposed to the data at bias point (ii), is shown in image (ii b). Here we assumed that the LTSLM signal is dominantly sensitive to the magnetic field $B$, i. e. $\Delta V \propto B^2$.
From the LTSLM data we infer $\lambda_x \approx \lambda_y \approx 100$ $\,\mu$m. Using $f=c(\lambda_x^{-2}+\lambda_y^{-2})^{1/2}$, with $f$ = 0.67 THz, we find a mode velocity $c \approx 4.7 \cdot 10^7$m/s, which is fully compatible with the \textit{in-phase} mode velocity at the elevated temperatures (60--70 K) estimated for the ´´cold'' part of the mesa \cite{Wang09a,Yurgens10}. Other (collective) mode velocities are by a factor of at least 2 lower. If the junctions would oscillate incoherently, the relevant mode velocity is the Swihart velocity $\bar{c} \approx 3 \cdot 10^5$m/s, which is even lower. We thus conclude that all junctions oscillate \textit{in-phase}.

\begin{figure}[tb]
\includegraphics[width=\columnwidth,clip]{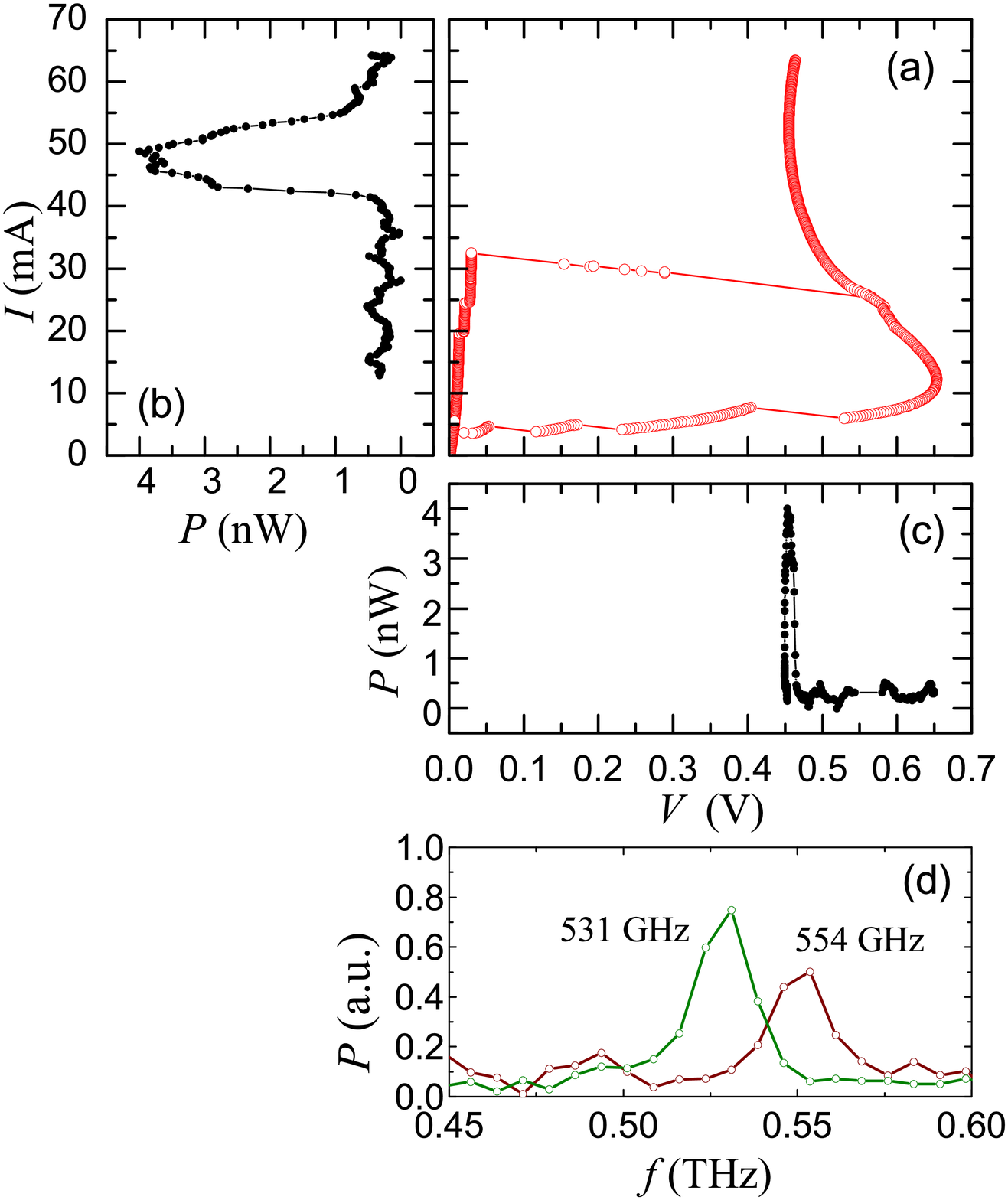}
\caption{(color online).
IVCs (a) sample 2 at $T$ = 40 K. Emission power detected by the bolometer as a function of current (b) and voltage (c). (d) shows two spectra of the emitted power at $V=$ 440 mV and $V=$ 451 mV.
}
\label{fig:4}
\end{figure}
%

%
\begin{figure}
\includegraphics[width=\columnwidth,clip]{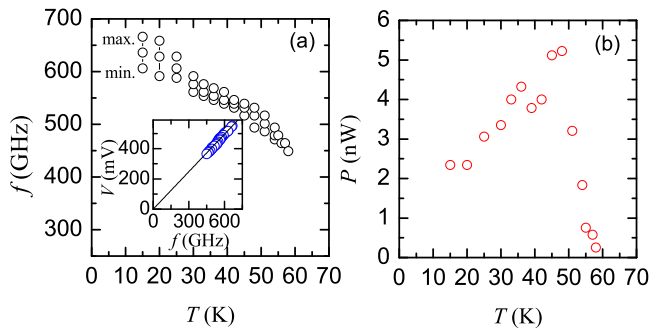}
\caption{(color online)
Range of emission frequencies $f$ (a) and detected emission power (b) of sample 2 as a function of bath temperature. Inset in (a) shows the voltage across the mesa as a function of $f$. Solid line in the inset is function $V=Nf\Phi_0$,  with $N$=400.
}
\label{fig:5}
\end{figure}

We next discuss emission data of sample 2.
IVCs, the bolometric response and Fourier spectrometer data are shown in Fig. \ref{fig:4}. The largest emission was detected near a voltage of 456 mV, c.f. Fig. \ref{fig:4} (b), (c). 
Fig. \ref{fig:4}(d) shows Fourier spectra at $V=$440 mV and $V=$451 mV. From the emission peaks at, respectively, 531 GHz and 545 GHz we find the Josephson relation $V/N =\Phi_0f$ to hold, with $N \approx 400$ (\ie somewhat less than the 470 junctions expected from the meas thickness). The 16 GHz linewidth is at the resolution limit.
LTSLM revealed that a hot spot formed for $I>25$mA, covering about half of the mesa area for the bias where maximum emission was detected. A standing wave pattern was vaguely visible, however further analysis of the LTSEM data was prevented by a spurious strong response, presumably caused by additional IJJs switching to the resistive state when illuminated by the laser.
Fig. \ref{fig:5} shows emission data taken at different bath temperatures. The frequency range of detectable emission is shown in Fig. \ref{fig:5} (a).
Radiation was found between 15 K and 58 K.  At the lower temperatures the emission frequencies varied continuously between 600 GHz and about 750 GHz; with increasing bath temperature the frequency of maximum emission power decreased, to about 450 GHz near 55 K; also, the frequency tunability decreased with increasing bath tempeature but still was on the order of 10$\%$ at 50 K. In all cases we found the Josephson relation $V/N =\Phi_0f$ to hold [c.f. inset of Fig. \ref{fig:5} (a)], with $N = 400$. Thus, in contrast to the ``low bias'' case, where the emission frequency is essentially fixed by the geometry of the mesa, it can be tuned in the hot spot regime. This can be understood if one accepts that one of the roles of the hot spot is to vary the length of the ``cold'' part of the mesa until, eventually, resonant conditions are achieved. Further, both with increasing dc input power and with increasing bath temperature, the (in-phase) mode velocity decreases, from about $7 \cdot 10^7$m/s for low $T$, towards zero when the cold part of the mesa reaches $T_c$ \cite{Wang09a}. Thus, the change of $f$ with bath temperature essentially reflects the decrease in mode velocity.

Still, the hot spot might even take an active role in THz generation and/or synchronization. For sample 1, the (resolution limited) 12 GHz linewidth of radiation is more than 50 times smaller than the oscillation frequency. For sample 2, $\Delta f/f < 0.03$ at $T$ = 40 K.
If, in the hot spot regime, $\Delta f$ were determined by the quality factor $Q$ of the cavity, one would require a $Q = f/\Delta f $ of at least 30--50, which is very unlikely to be the case at the elevated temperatures of the hot spot regime, where we expect $Q < 5$. Thus, although the  cavity seems to be a necessary ingredient for synchronization, the (nonequilibrium) phenomena, in the spirit of Refs. \cite{Lee00,
Kume99, Krasnov06, Krasnov09}, taking place at the hot spot edge may play an important role as well in synchronizing the junctions, eventually leading to the small values of $\Delta f$ observed. The fact that the Josephson frequency-voltage relation holds rules out a purely quasiparticle related process, however, should not contradict a scenario of interacting Josephson and quasiparticle currents.

In summary, we have investigated THz emissison from intrinsic Josephson junction stacks by off-chip emission measurements and by low temperature scanning laser imaging. The stacks emit radiation coherently at high input power where a hot spot has formed. Here, standing wave patterns are observed pointing to the importance of cavity resonances for synchronization. By changing the size of the ``cold'' part of the mesa the hot spot effectively adjusts the size of the cavity, making the emission frequency tunable. The linewidth of radiation, however, is too small to be determined by the quality factor of the cavity alone. An additional mechanism of synchronization seems to play a role and may arise from nonequilibrium processes at the edge of the hot spot.

We thank A. Yurgens, V.M.Krasnov, L.Ozyuzer, K. Kadowaki, I. Iguchi, K. Nakajima and C. Otani for valuable discussions.
Financial support by the strategic Japanese-German International Cooperative Program of
the JST and the DFG, and Grants-in-Aid for scientific research from JSPS is gratefully acknowledged.


%
%

\end{document}